\pdfoutput=1 
\documentclass[
  aps,%
 twocolumn,%
 groupedaddress,%
 superscriptaddress,%
  showpacs,%
 letterpaper,%
 amsfonts,%
 footinbib,%
 floatfix,%
]{revtex4-2}
 
\usepackage{dcolumn}
\usepackage{bm}

\usepackage{hyperref}
\hypersetup{
    colorlinks=true,
    citecolor=blue,
    linkcolor=blue , 
    filecolor=cyan,      
    urlcolor=magenta,
    pdfcreator = {\LaTeX\ and \flqq hyperref\frqq},
}

\RequirePackage{graphicx,amsmath,amssymb,bm}
\usepackage{bbm}
\usepackage{float}
\usepackage{framed} 
\usepackage{amsthm}
\usepackage{caption}
\usepackage{subcaption}
\usepackage[normalem]{ulem}
\usepackage{comment}
\usepackage{slashed} 
\usepackage{upgreek}
\usepackage{xcolor}
\usepackage{enumitem}

\graphicspath{ {./Figs/} }

\newcommand{\bb}{\mathbb}

\newcommand{\Qq}{{\cal{Q}}}

\newcommand{\bbz}{\bb{Z}}

\newcommand{\DKW}{D_{KW}}

\def\be{\begin{equation}}
\def\ee{\end{equation}} 
\def\bsh{\begin{shaded}}
\def\esh{\end{shaded}} 
\def\bpm{\begin{pmatrix}}
\def\epm{\end{pmatrix}}

\begin{document} 

\title{$\mathbb{Z}_2$ gauging and self-dualities of the $XX$ model and its cousins} 

\author{Lei Su} 
\affiliation{Department of Physics, University of Chicago, Chicago, Illinois 60637, USA
}

\begin{abstract} 
In this work, we investigate the one-dimensional $XX$ lattice model and its cousins through the lens of momentum and winding $U(1)$ symmetries. We distinguish two closely related $\mathbb{Z}_2$ symmetries based on their relation to the $U(1)$ symmetries, and establish a web of $\mathbb{Z}_2$-gauging relations among these models, rooted in two fundamental seeds: the $ XY \pm  YX$ models. These two seeds, each self-dual under gauging of the respective $\mathbb{Z}_2$-symmetries, possess manifestly symmetric conserved charges, making transparent the connection between the noninvertible symmetries and the Kramers-Wannier duality. By leveraging the self-dualities of these two seed models, we derive the self-dualities of their cousins, including the $XX$ model and the Levin-Gu model, through appropriate gauging procedures. Moreover, under these gauging schemes, the lattice T-duality matrices take the form of the identity matrix. These lattice models flow to the $c =1$ compact boson conformal field theory, with a twist that depends on the lattice size modulo four.  Finally, we unify the mapping structures of local conserved charges across these models, providing a comprehensive framework for understanding their symmetries and dualities.  
\end{abstract} 
\maketitle


\section{Introduction} 
The transverse field Ising model (TFIM) and the $XX$ model on a 1D lattice are well studied and serve as building blocks for many more complicated spin models. They can be mapped to free fermions and are thus exactly solvable. They are integrable and possess an infinite number of local conserved charges. In the infrared (IR), it is believed that the critical TFIM flows to the Ising conformal field theory (CFT) with central charge $c = 1/2$ while the $XX$ model flows to a compact boson CFT with $c = 1$ and radius $R = \sqrt{2}$ \cite{francesco2012conformal, ginsparg1988applied}. Recently, there has been a resurgence of research interests in both lattice models from the perspective of gauging, noninvertible symmetries, and dualities (see, e.g., Refs.~\cite{aasen2016topological, lootens2023dualities, li2023non, cheng2023lieb, seiberg2024majorana, seiberg2024non, okada2024non, pace2024lattice}).

Gauging, i.e., coupling the system to gauge fields, is a useful technique to detect and to construct noninvertible symmetries, i.e., symmetry operators without an inverse \cite{schafer2024ictp, shao2023}, and dualities.  The Kramers-Wannier (KW) duality in the critical Ising model provides the simplest example \cite{kramers1941}, arising from the self-duality under gauging of the spin-flip $\bbz_2$ symmetry. Whether a global symmetry can be gauged consistently determines whether the symmetry has a 't Hooft anomaly \cite{hooft1980naturalness}. Such anomalies constrain renormalization group flows and ground-state phases. 

The matching of 't Hooft anomalies on lattice models and their continuum limits is a fundamental problem in the study of quantum systems. Given a 't Hooft anomaly of some global symmetries in the IR, a key question arises: Are these symmetries emergent or exact? If they are exact, then how do the anomalies manifest in the ultraviolet (UV) or on the lattice? As an example, it is well known that the compact boson CFT exhibits a mixed anomaly between the momentum $U(1)$ symmetry and the winding $U(1)$ symmetry. Recently, Ref.~\cite{pace2024lattice} has identified both $U(1)$ symmetry generators in the $XX$ lattice model. Interestingly, while these generators commute in the IR, they form a (noncommutative) Onsager algebra on the lattice \cite{onsager1944}. Notably, the mixed anomaly between them is already visible on the lattice level. 

In this work, we study two classes of models unitarily equivalent to the $XX$ model with uniform and with staggered coupling constants, i.e., $ \sum_j^L (\pm 1)^j (X_j X_{j+1} +Y_j Y_{j+1})$, respectively. $X, Y, Z$ are the Pauli matrices. These two models are equivalent when the lattice size $L = 0 \mod 4$, but are different otherwise. They are also, respectively, equivalent to the $XY \pm YX$ models with Hamiltonian $\sum_j^L X_j Y_{j+1} \pm X_j Y_{j+1}$, and the $X \pm ZXZ$ models with Hamiltonian  $\sum_j^L X_j  \pm Z_{j-1} X_j Z_{j+1}$, for $L = 0 \mod 2$. The $X - ZXZ$ model is often called the Levin-Gu model \cite{levin2012braiding}. We use the term ``the $XX$ model and its cousins" to refer to these models collectively. 

In the following, we motivate the models through their relations to the local charges of the TFIM before delving into the discussion of two closely related yet subtly distinct $\bbz_2$ symmetries for one of the $U(1)$ symmetries and their gauging, which we refer to as $\bbz_2^{+}$-gauging and $\bbz_2^{-}$-gauging.   Using the $XY \pm YX$ models as foundational seeds, we establish a web of $\bbz_2$-gauging relations for the $XX$ model and its cousins, demonstrating that all these models exhibit self-duality under appropriately chosen gauging procedures. Under these gauging schemes, all lattice T-duality matrices assume the simple form of the identity matrix. These lattice models flow to the compact boson CFT, with a twist that depends on $L \mod 4$, analogous to the discussions in Ref.~\cite{cheng2023lieb}. The noninvertible symmetries arising from $\bbz_2^{\pm}$-gauging are directly linked to two types of KW dualities $D_{\pm}$. In this framework, we also present the operator algebras of the symmetries and relations of $U(1)$ charges in a manifestly symmetric representation. In the main text, we focus on the case where $L$ is even. A brief discussion for odd $L$ is relegated to Appendix~\ref{append:odd}.

\begin{figure*}
    \centering
    \includegraphics[width=0.98\linewidth]{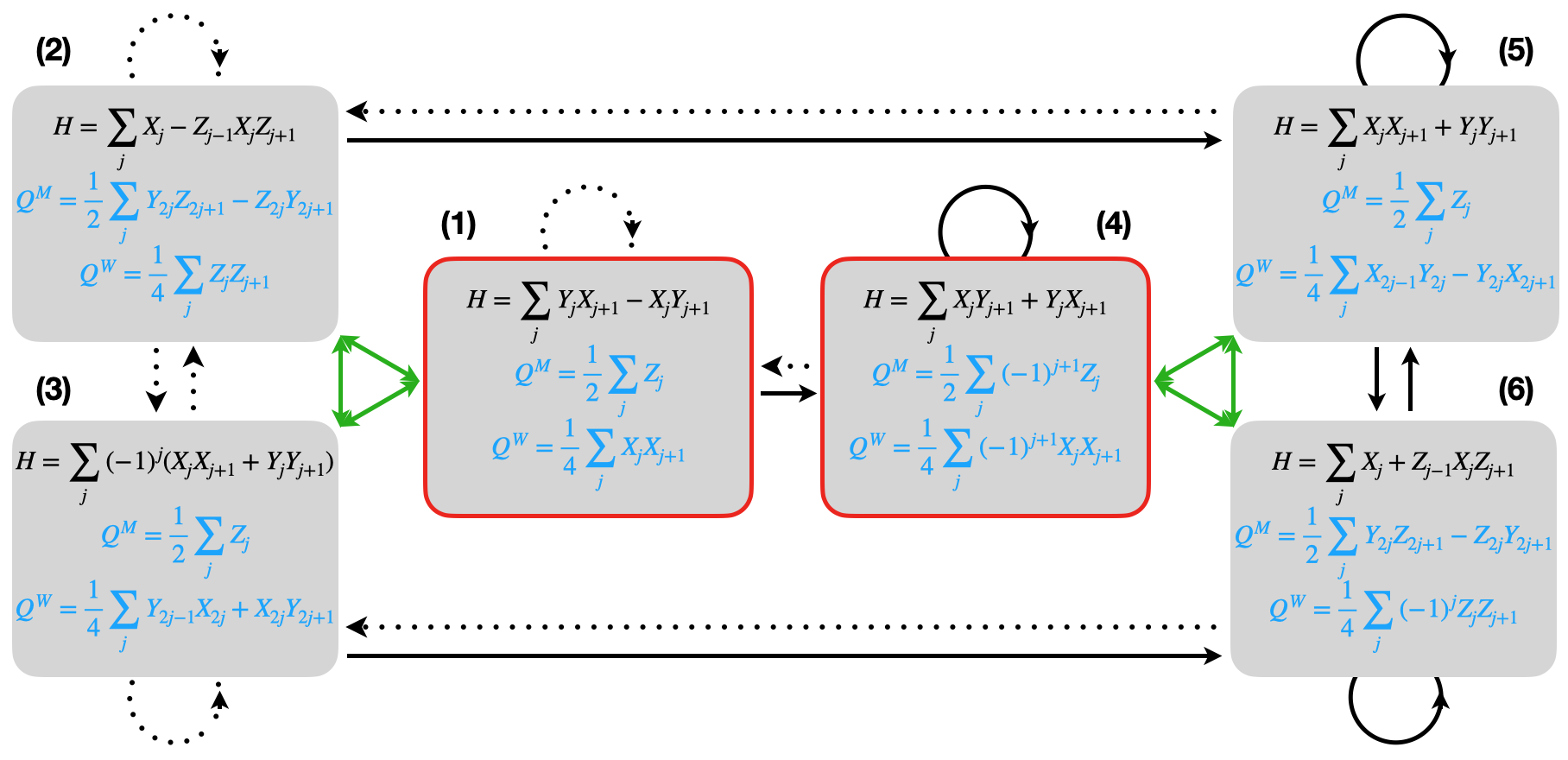}
    \caption{$\bbz_2$-gauging web of the $XX$ model and its five other cousins. The inner layer consists of two seeds at the center [(1) and (4), circled by red boundaries]: the $XY\pm YX$ model.  Dotted arrows and  solid arrows represent $\bbz_2^{+}$ gauging and $\bbz_2^{-}$ gauging, respectively. Green arrows represent unitary equivalences (for all even $L$), and connect the inner layer to the outer layer.  The three models on the left (right) are closed under $\bbz_2^{+}$-gauging ($\bbz_2^{-}$-gauging). The three models on the left (1-3) and the three models on the right (4-6) are related by $\bbz_2^{\mp}$-gauging. All models are self-dual under appropriate gauging procedures. Two quantized $U(1)$ charges,  $Q^M$ and $Q^W$, of each model are shown in the boxes.}
    \label{fig:1}
\end{figure*}

\section{Local charges of the TFIM}
Consider the TFIM in 1D
\be H_{TFIM}= J H_1 + h H_2, \ee with $ \quad H_1 =\sum_j Z_j,\ \ H_2 =\sum_j X_j X_{j+1}$. $J$ and $h$ are coupling constants.
We fix the lattice size $L$ to be even and impose periodic boundary conditions. The model has the famous KW duality $D_{\text{KW}}$ at $J =h$ that maps $Z_j \to X_{j} X_{j+1}$ and $X_j X_{j+1} \to  Z_{j+1}$.
The TFIM is integrable and has infinitely many  local conserved charges whose explicit form is known \cite{grady1982}. In particular, charges of the form
\begin{align}
  & q_{2r-1}= \\
   &\sum_j Y_{j} Z_{j+1}\cdots Z_{j+r-1} X_{j+r}- X_{j} Z_{j+1}\cdots Z_{j+r-1} Y_{j+r}  \nonumber 
\end{align}  
are independent of the coupling constant $J$ and $h$, which implies that $[q_{2r-1}, H_1] = [q_{2r-1}, H_2] = 0$. In addition, $q_{2r-1}$ commutes with each other for all integer $r$.

Since terms in $H_1$ commute with each other, it is easy to see that eigenvalues of $H_1$ take values in $2\bbz$ because $L$ is even. Similarly, $H_2$ is quantized. However, since the product of $X_j X_{j+1}$'s is 1, the eigenvalues of $H_2$, though quantized, is equal to $L \mod 4$.  Thus, $H_2/4$ is integer-quantized (half-integer-quantized) for $L = 0 \mod 4$ ($L = 2\mod 4$). This difference has important consequences in the IR \cite{cheng2023lieb}. 

As a result, if we treat $q_{2r-1}$ as a Hamiltonian, then it has at least two $U(1)$ symmetries generated by $H_1$ and $H_2$. Focusing on $r =1$, we can define the $XY -YX$ model \footnote{We follow the notation and conventions in Ref.~\cite{pace2024lattice} for the reader's convenience. The minus sign as compared to the name ``$XY -YX$'' is adopted for later convenience.}:
\be H_{XY -YX} = \sum_j Y_j X_{j+1} - X_j Y_{j+1},
\label{eq:xy}
\ee 
with $U(1)$ symmetries generated by 
\be 
Q^M = \frac{1}{2} \sum_j Z_j, \quad  Q^W =  \frac{1}{4} \sum_j X_j X_{j+1}.
\label{eq:qmqw}
\ee
Notably, we will see later that
$Q^M$ and $Q^W$ are exactly the charges generating, respectively, the momentum $U(1)^M$ symmetry and the winding $U(1)^W$ symmetry and that exchanging momentum and winding charges is directly related to the KW duality in the TFIM. 
Using the Jordan-Wigner transformation \cite{jordan1928}, we can map $H$ to the staggered fermion Hamiltonian (up to boundary conditions) as in Ref.~\cite{chatterjee2025quantized}, and map $Q^M$ and $Q^W$ to the vector charge $Q^V$ and the axial charge $Q^A$, respectively. The rest of the local charges $q_{2r-1}$ ($r>1$) are mapped to the symmetric deformations they considered. Commutative with $q_1$, these deformations $q_{2r-1}$ ($r>1$) do not lift the gaplessness once $U(1)^M$ and $U(1)^W$ are preserved. This is the ``symmetry-enforced gaplessness" discussed in Refs.~\cite{chatterjee2025quantized, pace2024lattice}.  Note that the $XY- YX$ model has many other quantized charges generated by $Q^M$ and $Q^W$ through pivoting \cite{tan2023pivot, jones2024pivoting}, which we will discuss later. It also has a charge conjugation symmetry $\bbz_2^C$ generated by $\prod_j X_{2j} Y_{2j+1}$, which flips $Q^M \to - Q^M$ and $Q^W \to -Q^W$.

By the same analysis, we can relate the $XY+YX$ model to the TFIM with staggered coupling constants.

\section{Two $\bbz_2$ symmetries}
The Hamiltonian in Eq.~(\ref{eq:xy}) has an obvious on-site $\bbz_2$ symmetry generated by
$\eta_Z \equiv \prod_j Z_j$. $\eta_Z$, however, is different from $\eta'_Z \equiv \prod_j (-1)^j Z_j$ (which in this case equal to $e^{i \pi Q^M}$) by staggered phase factors $\prod_j (-1)^j = (-1)^{L/2}$. When $L =0 \mod 4$, $\eta_Z = \eta'_Z$. However, when $L = 2 \mod 4$, $\eta_Z = -\eta'_Z$. The defining difference is that one generates the $\bbz_2$ subgroup of a $U(1)$ regardless of (even) $L \mod 4$ while the other does not. In the case of the $XY -YX$ model, it is $\eta'_Z$ that can be embedded into the $U(1)$ symmetry. However, we will see later that the roles of $\eta_Z$ and $\eta'_Z$ are switched in the $XY +YX$ model. This subtle difference leads to different $\bbz_2$-gauging. For convenience, we call it a $\bbz_2^-$-gauging if the corresponding $\bbz_2$ can be embedded in a $U(1)$ symmetry for all even $L$ as described above while the other a $\bbz_2^+$-gauging otherwise. It is known that gauging a discrete symmetry $\bbz_2$ generates a dual $\bbz_2$ symmetry which, if gauged, reproduces the original $\bbz_2$ symmetry \cite{Bhardwaj2024}. The dual $\bbz_2$ symmetry of $\bbz_2^{\pm}$ can be either of $\bbz_2^{\pm}$, depending on the system.  In fact, gauging these two $\bbz_2$ symmetries leads to a web of $\bbz_2$-gauging relations for the $XX$ model and its cousins (Fig.~\ref{fig:1}) we now discuss. 

\section{Web of $\bbz_2^{\pm}$-gauging}
There are many different ways to gauge an on-site $\bbz_2$ symmetry. A general gauging procedure in 1D is as follows. (i) Gauging: Place one gauge Ising spin, such as $\tilde{Z}_{j, j+1}$, on each bond of the lattice, couple the added gauge spins to neighboring spins on the sites of the original lattice properly, and enforce the gauge conditions, e.g.,
$ \tilde{X}_{j-1, j} (\pm 1)^j Z_j \tilde{X}_{j, j+1} =1$, on each site. (ii)  Disentangling: Apply a controlled-$Z$ transformation (or a similar operation) to each pair of neighboring spins to disentangle spins on the sites from spins on the bonds, and fix spins on the sites.  This procedure is far from unique: It can be proceeded by a unitary transformation on the sites and followed by another unitary transformation on spins on the bonds. A common practice is to sandwich the controlled-$Z$ transformation between Hadamard transformations (see Appendix \ref{Append:transf}) on every spin. More details can be found in Appendix \ref{append:gauging}. 
Following this procedure, we see that gauging the $\bbz_2^{\eta_Z}$ symmetry in Eq.~(\ref{eq:xy}) leads to itself. Thus it is self-dual under this $\bbz_2^+$-gauging. However, if we gauge $\bbz_2^{\eta'_Z}$, then we obtain the $XY + YX$ model (see Fig.~\ref{fig:1}) \footnote{Note that if we view the gauge conditions as emergent in the IR, then gauging a subgroup $\bbz_2 \subset U(1)$ without the disentangling step can give rise to an intrinsically gapless symmetry-protected topological phase if we preserve the quotient $U(1)/\bbz_2$ symmetry and the dual $\bbz_2$ symmetry \cite{su2024}.}:
\be H_{XY +YX}=  \sum_j X_j Y_{j+1} + Y_j X_{j+1},
\label{eq:xyyx}
\ee with integer-quantized conserved local $U(1)$ charges
\be 
\widetilde{Q}^{M} = \frac{1}{2} \sum_j (-1)^{j+1}  Z_j, \quad \widetilde{Q}^{W} =  \frac{1}{4} \sum_j (-1)^{j+1}  X_j X_{j+1}.
\label{eq:qmqwtil}
\ee 
Note that $\eta_Z = e^{i \pi \widetilde{Q}^M}$ in this case. Thus gauging $\bbz_2^{\eta_Z}$ is a $\bbz_2^-$-gauging while gauging $\bbz_2^{\eta'_Z}$ is a $\bbz_2^+$-gauging. For the $XY + YX$ model, the $\bbz_2^-$-gauging leads back to itself while the $\bbz_2^+$-gauging leads to the $XY - YX$ model. These relations are reflected in the web between the two seeds in the inner layer in Fig.~\ref{fig:1} \footnote{For the purpose of clarity, we do not study the gauging of $\bbz_2^C$ in this work.}.

It is easy to see that the $XY - YX$ model and the $XY + YX$ model are unitarily equivalent if $L = 0 \mod 4$ but are not equivalent if $L = 2 \mod 4$. In fact, the Hamiltonians of the $XY \pm YX$ model are unitarily equivalent to $ \sum_j (\pm 1)^j (X_j X_{j+1} +Y_j Y_{j+1})$ by performing $X_j \to Y_j$ and $Y_j \to \pm X_j$ on the odd sublattice, respectively. If $L = 2 \mod 4$, then $ \sum_j (- 1)^j (X_j X_{j+1} +Y_j Y_{j+1})$ can be viewed as the uniform $XX$ model with twisted boundary conditions.  By making use of controlled-$Z$ gates, we can see that they are also unitarily equivalent to $\sum_j X_j  \pm Z_{j-1} X_j Z_{j+1}$ (see Appendix \ref{append:web}). The unitary equivalences are represented by two triangles of green arrows in Fig.~\ref{fig:1}.  The corresponding $Q^M$'s and $Q^W$'s are shown in the boxes. 

Performing $\bbz_2^{+}$-gauging and $\bbz_2^-$-gauging produces the arrows in  the outer layer of the web (see Appendix \ref{append:web}). We notice that the three models on the left are closed under $\bbz_2^{+}$-gauging and the three models on the right are closed under $\bbz_2^{-}$-gauging. A freedom in composing a gauging operation with unitary transformations can, for example, modify the gauging map from box (2) to (3) in Fig.~\ref{fig:1} to the one from box (2) to itself. Performing $\bbz_2^{-}$-gauging ($\bbz_2^+$-gauging) of the three models on the left (right) leads to the three models on the right (left). Moreover, by choosing the gauging procedure appropriately, all six models are self-dual.

\section{Self-dualities}
\subsection{Self-duality of the $XY +YX$ model}
Let us first focus on the $XY+YX$ model in 
Eq.~(\ref{eq:xyyx}). It is believed to flow to the compact boson CFT for a compact scalar field $\phi$ (see Appendix \ref{append:cft}). Interpreting $X_j + i Y_j \propto  e^{\pm i\phi}$, we can view $\widetilde{Q}^M$ as the momentum $U(1)^M$ symmetry generator on the lattice level \footnote{The interpretation is most natural for the $XX$ model where $S_j^{\pm }= X_j \pm i Y_j \sim e^{\pm i\phi}$ for all $j$ and the Hamiltonian can be written as $H =\sum_j S_j^+ S_{j+1}^- +S_j^- S_{j+1}^+$.  The on-site $U(1)^M$ action rotates the phase $\phi$. Since other models are unitarily equivalent to the $XX$ model (up to twisted boundary conditions), we can identify the momentum charge in them.}. $\widetilde{Q}^W$, however, is interpreted as the winding $U(1)^W$ charge. Since $\eta_Z =e^{i\pi \widetilde{Q}^M}$, gauging  $\bbz_2^{\eta_Z}$ reduces $U(1)^M$ to $U(1)^M/\bbz_2$. $\widetilde{Q}^M$, which takes values in $\bbz$, now becomes $2\widetilde{Q}^W$, which takes values in $2\bbz$. Similarly, $U(1)^W$ is extended by $\bbz_2$, and the gauged winding charge takes values in $\bbz/2$.
Note that at this stage $2\widetilde{Q}^W$ and $\widetilde{Q}^M/2$ are still viewed as the gauged momentum charge and the gauged winding charge, respectively. The T-duality transformation is defined as an operation that switches the momentum charge and the winding charge. Under this gauging scheme, the T-duality transformation is an identity operation. Self-duality under gauging $\bbz_2^{\eta_Z}$  induces a noninvertible symmetry. It is easy to see that it is equivalent to a variant of a KW duality transformation $D_-$: $Z_j \to X_{j} X_{j+1} $ and $ X_j X_{j+1} \to  -Z_{j+1}$. It is a composition of the conventional KW $D_{\text{KW}}$ with $\eta_X \equiv \prod_j X_j$.   

The result is compatible with the continuum theory, the compact boson CFT with $c = 1$ at $R =\sqrt{2}$, which is self-dual under gauging $\bbz_2^-$ followed by a T-duality transformation: Gauging a $\bbz_2 \subset U(1)$, reduces $R \to R/2$, while T-duality maps $R \to 1/R$ by $\Qq^M \leftrightarrow \Qq^W$ (see Appendix \ref{append:cft}). Here, we used integer-valued $\Qq^M$ and $\Qq^W$ to denote the corresponding momentum and winding charges in the continuum theory. The spectrum is given by
\be 
E = \frac{1}{2} [\frac{(\Qq^M)^2}{R^2} + R^2 (\Qq^W)^2],
\label{eq:E}
\ee 
which is invariant under $\Qq^M \to R^2 \Qq^W$ and  $\Qq^W \to \Qq^M/R^2$.

\subsection{Self-duality of the $XY- YX$ model}
\label{subsect_XY}
In the $\bbz_2^{\eta_Z}$-gauging for the $XY+YX$ model described above, odd charges of $Q^M$ are projected out. For the $XY-YX$ model, $\eta_Z$ is not a subgroup of $U(1)^M$ in general. Gauging $\bbz_2^{\eta_Z}$ is thus a $\bbz_2^+$-gauging. It still maps $Q^M \to 2Q^W$ and $Q^W \to Q^M/2$.   Since $\eta_Z = (-1)^{L/2} \eta'_Z$, if $L = 0 \mod 4$, then odd charges are projected out. However, if $L =2 \mod 4$, then it is the even charges that are projected out. This observation is compatible with the fact that $Q^W$ is half-integer-quantized for $L = 2\mod 4$. If we instead gauge $\bbz_2^{\eta'_Z} \in U(1)^M$, then odd charges of $Q^M$ would be projected out regardless of $L$. Similar to $XY+YX$ model, gauging $\bbz_2^{\eta_Z}$ together with a lattice T-duality (which again takes the form of the identity matrix) leads to a self-duality of the $XY-YX$ model. The noninvertible symmetry operator in this case is simpler: It is exactly the KW duality transformation $D_+ =D_{\text{KW}}$:
$ Z_j \to X_{j} X_{j+1}, \quad  X_j X_{j+1} \to  Z_{j+1}$. 

The result is again compatible with the continuum limit, which is a compact boson CFT for $L = 0 \mod 4$ but twisted by $\eta_Z$ \cite{cheng2023lieb} for $L = 2\mod 4$. The spectrum is again given by Eq.~(\ref{eq:E}) except that now the eigenvalues of $\Qq^W$ are half-integers for $L = 2\mod 4$. This is a manifestation of the mixed anomaly between $U(1)^M$ and $U(1)^W$. The spectrum is invariant under the transformations $\Qq^M \to 2 \Qq^W$ and  $\Qq^W \to \Qq^M/2$ at $R =\sqrt{2}$. 

Note that the $XY- YX$ model also has a charge conjugation symmetry $\bbz_2^C$ that flips $Q^M \to - Q^M$ and $Q^W \to -Q^W$. Even though $Q^M$ and $Q^W$ do not commute \footnote{In fact, $Q^M$ and $Q^W$ commute within the ground state sub-Hilbert space, which is easy to check numerically for small lattice sizes, compatible with the conclusion that in the IR they become commutative \cite{pace2024lattice}.}, we can choose the states to be eigenstates of $Q^W$. Since the eigenvalues of $Q^W$ are half-integers, the ground state is at least doubly degenerate when $U(1)^W$ and $\bbz_2^C$ are preserved, compatible with the lattice calculation and the continuum limit for $L = 2\mod 4$ (see Appendix \ref{append:cft}) \cite{li2024decorated, cheng2023lieb}. 

Note that the discussion above can be generalized to odd $L$ in this case (see Appendix \ref{append:odd}).

\subsection{Self-duality of the $XX$ model}
Using the $XY \pm YX$ models as two foundational seeds, we can derive all the self-dualities in other models. As an example, we discuss the self-duality in the $XX$ model first \cite{verresen2021gapless}. In Ref.~\cite{pace2024lattice}, the authors adopted the conventional gauging procedure with controlled-$Z$ gates that maps the $XX$ model to the $X+ ZXZ $ model. The lattice T-duality matrix $U_T$, which is nontrivial, is easily constructed by composing the unitary equivalence from these models to the $XY+ YX$ model. The corresponding noninvertible symmetry in the $XX$ model has already been discussed there. The momentum charge and the winding charge are: $Q^M_{XX} =\sum_j Z_j /2$ and $Q^W_{XX} =\sum_j (X_{2j-1} Y_{2j} - Y_{2j}X_{2j+1})/4$.

However, as we emphasized, the gauging procedure is far from unique. We are allowed to perform a unitary transformation on the original spins before mixing them with gauged spins by controlled-$Z$ (or similar) gates, and a unitary transformation on the gauge spins afterwards. $U_T$ applied afterwards is simply such a unitary transformation. Making use of the self-duality of the $XY+ YX$ model and composing it with the unitary equivalence between it and the $XX$ model, we can also determine a gauging procedure that makes the self-duality of the $XX$ model more manifest. This gauging procedure is detailed in Appendix \ref{append:outerlayer}. In particular, under this gauging, $Z_{2j-1}\to X_{2j-1} Y_{2j}$, $Z_{2j}\to -Y_{2j}X_{2j+1}$, $X_{2j-1} Y_{2j} \to Z_{2j}$, and $Y_{2j}X_{2j+1}\to -Z_{2j+1}$.
Thus the self-duality maps $Q^M_{XX} \to 2Q^W_{XX}$ and $Q^W_{XX} \to Q^M_{XX}/2$. The lattice T-duality is the identity transformation.

\subsection{Self-duality of the Levin-Gu model}
We discuss one more example: the $X - ZXZ$ model, also known as the Levin-Gu model, whose Hamiltonian is   
\be 
H_{LG} = \sum_j X_j  - Z_{j-1} X_j Z_{j+1}.
\ee 
This model was proposed to describe the edge of a 2D topological phase protected by a $\bbz_2$ symmetry \cite{levin2012braiding}. In fact, it has (at least) two $U(1)$ symmetries generated by $Q^M_{LG}= \sum_j (Y_{2j} Z_{2j+1} - Z_{2j}Y_{2j+1})/2$ and $Q^W_{LG} = \sum_j Z_j Z_{j+1}/4$. 
Note that 
$ \eta_X = \prod_j X_j = e^{i\pi Q^M_{LG}}$ is on-site and can be gauged easily.  Gauging $\bbz_2^{\eta_X}$ using controlled-$Z$ gates produces the $XX$ model, and gauging $\eta'_X = \prod_j (-1)^j X_j$ naturally produces the $XX$ model with staggered couplings (see Fig.~\ref{fig:1}), which is unitarily equivalent to the Levin-Gu model. 

Similar to what we did for the $XX$ model, we can choose a gauging procedure by composing the unitary transformation between the Levin-Gu model and the $XY-YX$ model and the self-duality $\bbz_2$-gauging in the $XY-YX$ model. The gauged Hamiltonian is 
\begin{align}
        &  H'_{LG} =  \sum_j X_{2j} \tilde{Z}_{2j, 2j+1} +\tilde{Z}_{2j, 2j+1} X_{2j+1} \\
        & -Z_{2j-1} \tilde{Z}_{2j-1, 2j}  X_{2j} Z_{2j+1} -Z_{2j}X_{2j+1} \tilde{Z}_{2j+1, 2j+2}  Z_{2j+2},\nonumber
\end{align} 
 and the gauge conditions are
 \begin{subequations}
 \begin{align}
     \tilde{X}_{2j-1,2j} Y_{2j} \tilde{X}_{2j, 2j+1} Z_{2j+1} &=1, \\
      - Z_{2j} \tilde{X}_{2j, 2j+1} Y_{2j+1} \tilde{X}_{2j+1, 2j+2} &=1, 
 \end{align}
 \end{subequations}
 for all $j$. The disentangling transformation is given in Appendix \ref{append:outerlayer}. We can show that the Levin-Gu model is self-dual under this unconventional gauging procedure and that it maps $Q^M_{LG} \to 2Q^W_{LG}$ and $Q^W_{LG} \to Q^M_{LG}/2$. Note that a more
 conventional way of gauging $\eta_X$ via minimal coupling can also work, provided that a suitable unitary transformation on the bonds is chosen to follow the controlled-$Z$ transformation (or a similar operation) at the disentangling step (by chasing the diagram in Fig.~\ref{fig:1}).

\section{Noninvertible symmetries} 
Having shown that all six models are self-dual under appropriate gauging schemes, we now turn to the corresponding noninvertible symmetries. Under appropriate gauging schemes, all these symmetry operators map $Q^M \to 2Q^W$ and $Q^W\to Q^M/2$. Here, we use $Q^M$ and $Q^W$ for all models. As we have already mentioned, the noninvertible symmetry operator of the $XY-YX$ model is  
$D_+ = \DKW$ and the the noninvertible symmetry operator of the $XY+YX$ model is $D_- =  \eta_X \DKW$. The corresponding noninvertible symmetry operators for other models are obtained by unitary transformations and are described in Appendix \ref{append_noninv}.  

Note that other than the mixed anomaly between $U(1)$ symmetries and the type-III anomaly associated with two $U(1)$ symmetries and charge conjugation $\bbz_2^C$ \cite{pace2024lattice}, the noninvertible symmetry generated by $D_+ = \DKW$ can also be viewed as being anomalous \cite{seiberg2024non} regardless of $L \mod 4$, thus ruling out the possibility of a trivially gapped ground state. Since the $XY+YX$ model is obtained from the $XY-YX$ model by twisting the boundary with a discrete $\bbz_2$ defect, we expect the symmetry generated by $D_-$ to be anomalous for all even $L$ as well. By unitary equivalences, all noninvertible symmetries associated with the self-dualities in the six models should be anomalous.

We can also derive the operator algebras associated with the noninvertible symmetry generators. The algebras for the $XY\pm YX$ model follow directly from the operator algebra associated with $\DKW$ \cite{seiberg2024non}. For $D_+$ 
\begin{align}
    & \eta_Z^2 = 1, & \eta_Z D_+= D_+ \eta_Z = D_+, \nonumber \\
    & D_+^2 = (1 +\eta_Z) T , & T D_+ = D_+T =D_+^{\dagger},  
\end{align}
where $T$ is the translation $j \to j+1$ \footnote{Note that, as argued in Ref.~\cite{cheng2023lieb}, in the continuum limit, $Q^M$ and $Q^W$ flow to $\Qq^M$ and $\Qq^W$ respectively, but $T$ can flow to a nontrivial ``emanant'' internal symmetry.}. For $D_-$,
\begin{align}
    & \eta_Z^2 = 1, & \eta_Z D_-= D_- \eta_Z = D_-, \nonumber \\
    & D_-^2 = \eta_C (1 +\eta_Z) T, & T D_- = D_-T =D_-^{\dagger}, 
\end{align}
where $\eta_C= (-i)^{L/2} \prod_j X_{2j} Y_{2j+1}$ is the charge conjugation operator. 

\begin{figure}
    \centering
    \includegraphics[width=0.99\linewidth]{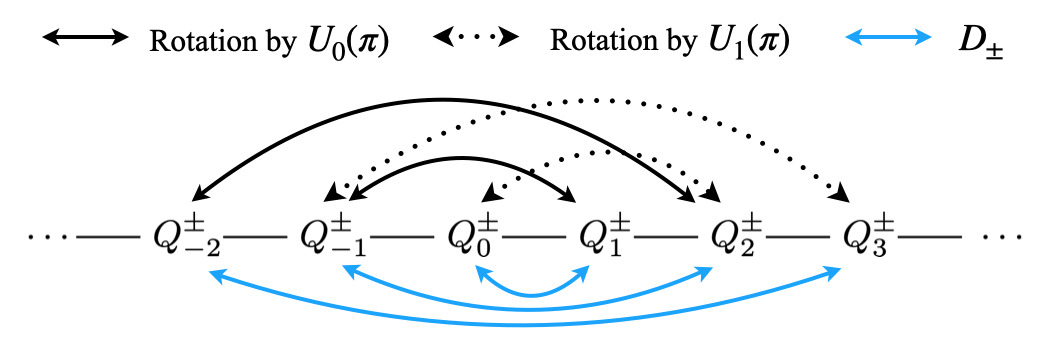}
    \caption{Maps between local charges $Q_n^{\pm}$. $U_0(\pi) = e^{i\pi Q_0^{+}/4}$ and $U_1(\pi) = e^{i\pi Q_1^{+}/4}$. $D_{\pm}$ relating $Q_n^{+}$'s to $Q_n^-$'s (up to unitary transformations) are not shown.}
    \label{fig:2}
\end{figure}

\section{Maps between local charges}
As already mentioned, the $XY-YX$ model has infinitely many quantized conserved local charges. $Q^M$ and $Q^W$ are just two of them. For example, by symmetry, $\sum_j Y_j Y_{j+1}/4$ is also a local charge, a special case of $Q^W$ rotated by $e^{i\theta Q^M}$. $\bbz_2^+$-gauging this charge yields another new charge $-\sum_j X_{j-1} Z_j X_{j+1}/2$. This process can be repeated indefinitely. The (unnormalized) charges are
\begin{align*}
  Q_n^{+} =\begin{cases}  \sum_j Z_j , \quad &n =0; \\
  (-1)^{n-1}\sum_j X_j Z_{j+1}\cdots Z_{j+n-1} X_{j+n}, \quad &n >0; \\
  (-1)^{n-1} \sum_j Y_j Z_{j+1}\cdots Z_{j-n-1} Y_{j-n}, \quad &n<0; 
  \end{cases}
\end{align*}
where $Q_0^+ = 2 Q^M$ and $Q_1^+ = 4 Q^W$ satisfy the Dolan-Grady relations \cite{dolan1982, vernier2019onsager}. Other $Q_n^+$'s, often called generalized cluster Hamiltonians \cite{verresen2017}, can be generated by rotating (also called pivoting) using $e^{i\pi Q_0^{+}/4}$ and $e^{i\pi Q_1^{+}/4}$. The maps between them have already been discussed in Refs.~\cite{tan2023pivot,jones2024pivoting} and are shown in Fig.~\ref{fig:2}. The corresponding charges of the $XY+YX$ model are similar \footnote{These charges are related to those obtained in Ref.~\cite{pace2024lattice} by a unitary transformation. They are manifestly more symmetric in this representation.}:  
\begin{align*}
  Q_n^{-} =\begin{cases}  \sum_j (-1)^{j+1}Z_j , \quad &n =0; \\
  \sum_j(-1)^{j+n} X_j Z_{j+1}\cdots Z_{j+n-1} X_{j+n}, \quad &n>0; \\
  \sum_j (-1)^{j} Y_j Z_{j+1}\cdots Z_{j-n-1} Y_{j-n}, \quad &n<0. 
  \end{cases}
\end{align*}
The difference is that the gauging is a $\bbz_2^-$-gauging and the noninvertible operator is $D_-$. The maps between charges by rotations (using $Q_0^-$ and $Q_1^-$) and the KW transformation are also shown in Fig.~\ref{fig:2}, which are isomorphic to those obtained in Ref.~\cite{pace2024lattice}. By unitary equivalence, the same mapping structure applies to other models.

We make a few comments here. 
First, even though $e^{i \theta Q_n^{\pm}/2}$  for $n=2k \in 2\bbz$ is non-on-site for generic $\theta$, 
$ e^{i \pi Q^{\pm}_{2k}/2}$ is equal to $\eta_Z$ or $\eta'_Z$ for all $n$. In other words, the $U(1)$ symmetries generated by $Q^{\pm}_{2k}$'s share a common $\bbz_2$ subgroup. Thus the dichotomy of $\bbz_2^{\pm}$-gauging applies to all $U(1)$ symmetries generated by $Q^{\pm}_{2k}$'s. In fact, $Q^{\pm}_{2k}/2$ all flow to $\Qq^M$ (and $Q^{\pm}_{2k-1}/4$ all flow to $\Qq^W$) in the IR \cite{pace2024lattice}. Even though we call $Q^M$ and $Q^W$ \textit{the} momentum and winding charges, momentum and winding charges have infinite lattice realizations. Second, there are many other local charges that are not shown, e.g., $q_{2r-1}$'s with $r>1$ for the $XY-YX$ model, which are invariant under the action of $Q_0^{+}$, $Q_1^{+}$, and $D_+$. Third, $\bbz_2^{\pm}$-gauging connects $Q_n^{+}$'s and $Q_n^{-}$'s but they are not shown in Fig.~\ref{fig:2}.

\section{Conclusion}
In this work, we studied two types of $\bbz_2$-gauging, namely, $\bbz_2^{\pm}$-gauging, and self-dualities in the $XX$ lattice model and its cousins. We highlighted the $XY\pm YX$ models, as they can be viewed as local charges of the TFIM, and used them as fundamental seeds to establish a web of $\bbz_2$-gauging relations among the related models. Exploiting the self-dualities of these two seeds and the unitary equivalences between them and other models, we determined appropriate $\bbz_2$-gauging procedures that render all models self-dual. Furthermore, we found that the lattice T-duality matrices take the form of the identity matrix.  These lattice models flow to the compact boson CFT, with a twist that depends on the lattice size $L \mod 4$ (see also Appendix~\ref{append:odd}).  We also demonstrated the direct connection of the noninvertible symmetries to the (anomalous) KW duality, obtaining their associated algebras in the $XY\pm YX$ models. Finally, we identified manifestly symmetric local charges and unified their mapping structures. Our work provides a comprehensive framework for understanding symmetries and dualities in the $XX$ model and its cousins.

We mention two directions for further exploration. First, it is known that gauging the charge conjugation $\bbz_2^C$ in the $XX$ model yields two decoupled critical Ising models. As discussed, the $XY\pm YX$ models are directly related to the TFIM. Thus, gauging $\bbz_2^C$ not only extends the web in Fig.~\ref{fig:1}, but also directly links the self-dualities of this family of models to those of two copies of the critical Ising model \cite{thorngren2024fusion, thorngren2024fusion2, choi2024self, perez2024notes, diatlyk2024gauging}, which should be elaborated. Second, generic gauging procedures induced by unitary transformations, as exemplified by the unconventional gauging for the Levin-Gu model, warrant further investigation.

\section{Acknowledgment} 
I thank Aashish Clerk, Ivar Martin, and Meng Zeng for discussion. This work was supported by the US Department of Energy, Office of Science, Basic Energy Sciences, Materials Sciences and Engineering Division.

\newpage
\appendix
\section{Useful transformations}
\label{Append:transf}
We list some transformations we use in this work here. 

(a) 
\textbf{Two-qubit controlled-$Z$ gate}: $CZ$. It is defined as
    \be 
    CZ_{ij} = \frac{1+ Z_i +Z_j -Z_i Z_j}{2}.
    \label{czdef}
    \ee 
    It changes the phase of the wavefunction by $-1$ if both spins are down. (Here, we identify a spin-1/2 with a qubit.) It is Hermitian and has order 2. In particular,
    \begin{align}
        CZ_{ij} X_i CZ_{ij} = X_i Z_j, & \quad CZ_{ij} X_j CZ_{ij} = Z_i X_j,   \\
         CZ_{ij} Y_i CZ_{ij} = Y_i Z_j,  & \quad  CZ_{ij} Y_j CZ_{ij} = Z_i Y_j   \\
          CZ_{ij} Z_i CZ_{ij} = Z_i, &\quad CZ_{ij} Z_j CZ_{ij} = Z_j.
    \end{align}
    For a periodic lattice of qubits, one can define a SPT entangler 
    $U = \prod_j^L CZ_{j, j+1}$ with the following properties:
    \begin{align}
        &  U X_j U^{-1} = Z_{j-1}X_j Z_{j+1},  \\ 
        & U  Z_{j-1}X_j Z_{j+1} U^{-1} =  X_j .
    \end{align}
    For even $L$, using the definition for $CZ$ in Eq.~(\ref{czdef}), we can see easily 
    \be 
    U = e^{i \pi \frac{1}{4}\sum_j^L(-1)^j Z_j Z_{j+1}}.
    \ee 
    Thus, $\frac{1}{4}\sum_j^L(-1)^j Z_j Z_{j+1}$ is a pivot Hamiltonian for $U$ \cite{tan2023pivot}.

     (b)
  \textbf{ Rotation of Pauli matrices}: \be  R_{\hat{n}}(\theta) = e^{-i \frac{\theta}{2} (\hat{n}\cdot \vec{\sigma})},
  \ee
  where $\hat{n}$ is a unit vector specifying the rotational axis, and $\vec{\sigma} =(X, Y, Z)$. It can be written as  
  \be 
  R_{\hat{n}}(\theta) = \cos \frac{\theta}{2} I - i \sin \frac{\theta}{2} \hat{n}\cdot \vec{\sigma}.
  \ee The rotational operator acts on the Pauli matrices as 
  \begin{align}
    & R_{\hat{n}}(-\theta) \vec{\sigma}  R_{\hat{n}}(\theta) = \nonumber \\
    & \cos \theta\  \vec{\sigma} + \sin \theta\  \hat{n}\times \vec{\sigma} + (1 -\cos \theta) (\hat{n}\cdot \vec{\sigma})\hat{n}.   
  \end{align}
  In particular,
  \be
  R_Z(-\theta) \bpm X \\ Y \epm   R_Z(\theta)= \bpm\cos \theta, -\sin \theta \\ \sin \theta,\ \ \cos \theta \epm \bpm X \\ Y \epm.
  \ee 
  The transformation holds for any cyclic permutation in the Pauli matrices.
  In this work, we frequently make use of special cases (up to cyclic permutations in the Pauli matrices).
    \begin{itemize}
    \item $X \to -Y$ and $Y \to X$ : $R_Z(\pi/2)$.
    \item $X \to -X$ and $Y \to -Y$: $R_Z(\pi) =-i Z$.
    \item $X \leftrightarrow Z$ and $Y \to - Y$ (Hadamard gate): $H_Y =i R_X(\pi)R_Y(\pi/2) =XR_Y(\pi/2) $.
    \item $X \to Y \to Z \to X$: $R_{\hat{n}_0}(2\pi/3)$ with $\hat{n}_0 = -(1, 1, 1)/\sqrt{3}$. Explicitly,
    \be 
 R_{\hat{n}_0}(2\pi/3) 
 =\frac{1}{2}[I + i(X+Y+Z)].\nonumber
 \ee
 It can also be decomposed into two simpler rotations, e.g., $R_Y(-\pi/2) R_Z(-\pi/2)$.
    \end{itemize}

\section{Gauging procedure}
\label{append:gauging}
In this appendix, we provide more details about the gauging procedure. To be more concrete, we use the $XY -YX$ model as an example.
\be H = \sum_j Y_j X_{j+1} - X_j Y_{j+1},
\label{eq:append_xy}
\ee 
and gauge $\eta_Z = \prod_j Z_j$.

(i) Gauging: Place one gauge Ising spin on each bond of the lattice, couple the added gauge spins to neighboring spins on the sites of the original lattice properly, and enforce gauge conditions on each site.

We are free to choose the Ising spin in the Bloch sphere based on $\tilde{X}$, $\tilde{Y}$, $\tilde{Z}$. To be concrete, we choose $\tilde{Z}_{j, j+1}$. Typical gauge constraints used are
\be  \tilde{X}_{j-1, j} Z_j \tilde{X}_{j, j+1} =1.\ee
 The Hamiltonian becomes 
\be H' = \sum_j  Y_j \tilde{Z}_{j, j+1} X_{j+1} -X_j \tilde{Z}_{j, j+1}Y_{j+1}.
\ee 
At this step, it also common to perform Hadamard transformations on lattice sites $X_j \leftrightarrow Z_j$ first to set the stage for the application of controlled-$Z$ gates in the next step. More generically, we can perform other (not necessarily on-site) unitary transformations on the spins on the original lattice.

Note that if we are to gauge $\eta'_Z =\prod_j (-1)^j Z_j$, we can impose the gauge condition
\be  \tilde{X}_{j-1, j} (-1)^j Z_j \tilde{X}_{j, j+1} =1\ee
for all $j$.

(ii) Disentangling: Apply a controlled-$Z$ transformation (or a similar operation) to each pair of neighboring spins to disentangle spins on the sites from spins on the bonds, and fix spins on the sites. As mentioned above, it is common to apply the Hadamard transformation on every spin first before applying controlled-$Z$ gates (see Appendix \ref{Append:transf}) on every pair of neighboring spins. After that, we can use the Hadamard transformation on every spin to rotate the basis back. (We call the composite transformation a Hadamard-rotated controlled-$Z$ transformation for simplicity and use $H (CZ) H$ to represent it.)  A Hadamard-rotated controlled-$Z$ transformation acts as 
\be 
     Z_j\to   \tilde{X}_{j-1,j} Z_j \tilde{X}_{j,j+1},\quad X_j \tilde{Z}_{j,j+1} X_{j+1}  \to \tilde{Z}_{j,j+1}.  
\ee
Under this unitary transformation, $H'$ is mapped to 
\begin{align}
    \tilde{H}& =  \sum_j  (-i)\tilde{X}_{j-1 j} Z_j \tilde{X}_{j, j+1}  \tilde{Z}_{j, j+1}  \\
    &- i  \tilde{Z}_{j, j+1} \tilde{X}_{j, j+1}Z_{j+1}\tilde{X}_{j+1, j+2} \nonumber \\
    & =  \sum_j  -\tilde{X}_{j-1 j} Z_j  \tilde{Y}_{j, j+1}+ \tilde{Y}_{j, j+1} Z_{j+1}\tilde{X}_{j+1, j+2}\nonumber
\end{align}
and 
the gauge condition is mapped to 
\be 
Z_j =1.
\ee 
Thus the original spins on the sites are totally disentangled from the gauge spins on the bonds. Fixing the gauge and plugging in $Z_j =1$, we obtain 
\be \tilde{H} =  \sum_j  \tilde{Y}_{j, j+1}  \tilde{X}_{j+1, j+2} -\tilde{X}_{j-1 j}   \tilde{Y}_{j, j+1}.
\ee 
Dropping the tildes and replacing $(j, j+1) \to j+1$, we can see that $\tilde{H}$ is identical to $H$ in Eq.~(\ref{eq:append_xy}).

Note that this process can be followed by any other unitary transformation on gauge spins on the bonds.

\section{Details for the main web}
\label{append:web}
In this appendix, we show the details of unitary transformations, gauging procedures for both the inner layer ($XY\pm YX$ models) and the outer layer, and noninvertible symmetry operations associated with self-dual $\bbz_2$-gauging in Fig.~\ref{fig:1}. 
\subsection{Unitary transformations in the web}
\begin{enumerate}[label=(\alph*)]
\item $2 \to 1$ 

$CZ$ on $(2j, 2j+1)$ followed by $U_{2 \to 1}$ where $U_{2 \to 1}$ is defined to be $X_j \to Y_j \to Z_j \to X_j$ on every site and $Y_j \to - Y_j$ and $Z_j \to - Z_j$ on odd sites.

\item $1 \to 3$  

$U_{1 \to 3}$: $X_j \to  Y_j$ and $Y_j \to -X_j$ on odd sites.

\item $2 \to 3$  

$CZ$ on $(2j, 2j+1)$ followed by $U_{2 \to 3}$ where $U_{2 \to 3}$ is defined to be 
 $X_j \to Y_j \to Z_j \to X_j$ on even sites and $Y_j \to - Z_j$ and $Z_j \to Y_j$ on odd sites.

\item $6 \to 4$

$CZ$ on $(2j, 2j+1)$ followed by $U_{6 \to 4}$ where $U_{6 \to 4}$ is defined to be $X_j \to Y_j \to Z_j \to X_j$ on every site followed by  $Y_j \to  -Y_j$ and $Z_j \to -Z_j$ on odd sites, and $X_j \to  -X_j$ and $Z_j \to -Z_j$ on even sites. 
\item $4 \to 5$

$U_{4 \to 5}$: 
$X_j \to  Y_j$ and $Y_j \to X_j$ on even sites.

\item $6 \to 5$ 

$CZ$ on $(2j, 2j+1)$ followed by $U_{6 \to 5}$ where $U_{6 \to 5}$ is defined to be $X_j \to Y_j \to Z_j \to X_j$ followed by  $Y_j \to  -Y_j$ and $Z_j \to -Z_j$ on odd sites, and  $Y_j \to  Z_j$ and $Z_j \to -Y_j$ on even sites.
\end{enumerate}

\subsection{Gauging in the web: Inner layer}
\begin{enumerate}[label=(\alph*)]
\item $1 \leftrightarrow 1$ 
  \begin{align*}
        & Z_j \to  \tilde{X}_{j-1,j}Z_j \tilde{X}_{j, j+1}, \quad \tilde{X}_{j,j+1} \to \tilde{X}_{j,j+1}, \nonumber\\
        & X_j \to X_j, \quad\quad \tilde{Z}_{j, j+1} \to X_{j} \tilde{Z}_{j, j+1} X_{j+1}.  
    \end{align*}

\item $4 \leftrightarrow 4$ 
    \begin{align*}
        & Z_j \to \tilde{X}_{j-1,j}Z_j \tilde{X}_{j, j+1}, \quad \tilde{X}_{j,j+1} \to \tilde{X}_{j,j+1}, \nonumber\\
        & X_j \to X_j, \quad\quad \tilde{Z}_{j, j+1} \to X_{j} \tilde{Z}_{j, j+1} X_{j+1}.  
    \end{align*} 
    followed by $Y_j \to -Y_j$ and $Z_j \to - Z_j$ on every site after dropping the tildes and replacing $(j, j+1) \to j+1$.
 \item $1 \to 4$  
    \begin{align*}
        & Z_j \to (-1)^j\tilde{X}_{j-1,j}Z_j \tilde{X}_{j, j+1}, \quad \tilde{X}_{j,j+1} \to \tilde{X}_{j,j+1}, \nonumber\\
        & X_j \to X_j, \quad\quad \tilde{Z}_{j, j+1} \to X_{j} \tilde{Z}_{j, j+1} X_{j+1},   
    \end{align*} 
    followed by $Z_j \to -Z_j$ and $X_j \to -X_j$ on even sites after dropping the tildes and replacing $(j, j+1) \to j+1$.  

\item $4 \to 1$.
    \begin{align*}
        & Z_j \to (-1)^j \tilde{X}_{j-1,j}Z_j \tilde{X}_{j, j+1}, \quad \tilde{X}_{j,j+1} \to \tilde{X}_{j,j+1}, \nonumber\\
        & X_j \to X_j, \quad\quad \tilde{Z}_{j, j+1} \to X_{j} \tilde{Z}_{j, j+1} X_{j+1},
    \end{align*} 
    followed by $Z_j \to -Z_j$ and $X_j \to -X_j$ on even sites after dropping the tildes and replacing $(j, j+1) \to j +1$.
    \end{enumerate}

    \subsection{Gauging in the web: Outer 
layer}
    \label{append:outerlayer}
Using the unitary transformations and the gauging procedure mentioned above, we can determine the following self-dual gauging procedures for the models in the outer layer by composing a unitary transformation on spins on the sites (original lattice), the gauging procedure for the $XY \pm YX$ models, and another unitary transformation on gauge spins on the bonds (dual lattice). Note that self-dual gauging procedures are not unique.
    \begin{enumerate}[label=(\alph*)]
    \begin{figure}[!htb]
    \centering
\includegraphics[width=0.9\linewidth]{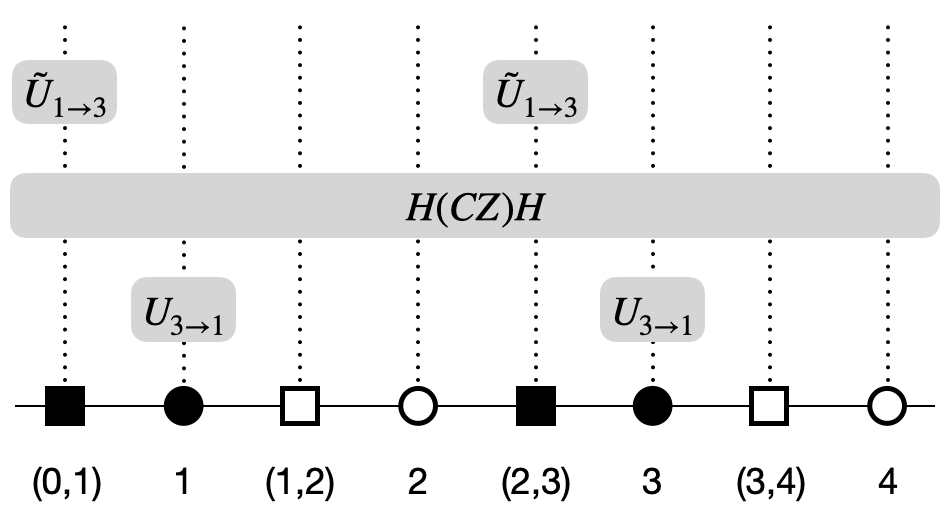}
    \caption{Self-dual disentangling transformation for $3 \leftrightarrow 3$. Circled (squared) sites represent the original (dual) lattice. Filled (empty) marks represent odd (even) sites or bonds. Operators with tildes act on the dual lattice. }
    \label{fig:3}
\end{figure}

    \item $3 \leftrightarrow 3$

    The gauged Hamiltonian is 
   \be H' = \sum_j (-1)^j (X_j \tilde{Z}_{j, j+1} X_{j+1} +Y_j \tilde{Z}_{j, j+1}Y_{j+1}).\ee
   The gauge condition is $\tilde{X}_{j-1,j}Z_j \tilde{X}_{j, j+1} =1$ for all $j$. The disentangling transformation is shown in Fig.~\ref{fig:3}.
    
    For odd $j$,
    \begin{align*}
        & Z_j \to \tilde{Y}_{j-1,j}Z_j \tilde{X}_{j, j+1}, \quad X_j \to -\tilde{Y}_{j-1,j}Y_j \tilde{X}_{j, j+1},  \\
        & \tilde{Z}_{j-1, j} \to X_{j-1} \tilde{Z}_{j-1, j} X_{j}, \quad  \tilde{X}_{j-1,j} \to \tilde{Y}_{j-1,j}. 
    \end{align*}  
    For even $j$,
\begin{align*}
        & Z_j \to \tilde{X}_{j-1,j}Z_j \tilde{Y}_{j, j+1}, \quad X_j \to X_j,  \\
        & \tilde{Z}_{j-1, j} \to X_{j-1} \tilde{Z}_{j-1, j} X_{j}, \quad  \tilde{X}_{j-1,j} \to \tilde{X}_{j-1,j}. 
    \end{align*}

 \begin{figure}[b]
    \centering
\includegraphics[width=0.9\linewidth]{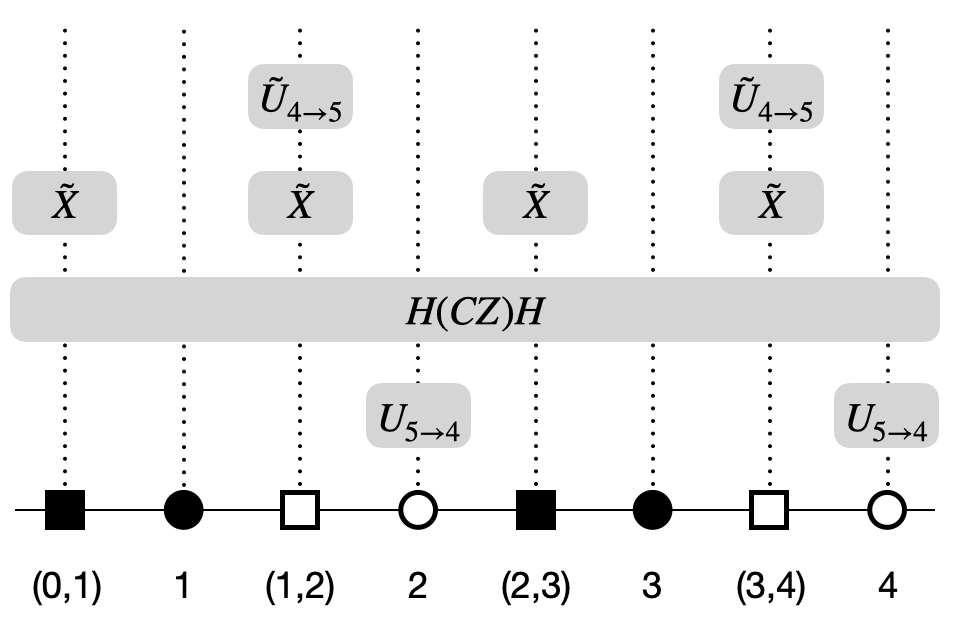}
    \caption{Self-dual disentangling transformation for $5\leftrightarrow 5$. Circled (squared) sites represent the original (dual) lattice. Filled (empty) marks represent odd (even) sites or bonds). Operators with tildes act on the dual lattice.}
    \label{fig:4}
\end{figure}
        
    \item $5 \leftrightarrow 5$

    The gauged Hamiltonian is 
   \be H' = \sum_j   X_j \tilde{Z}_{j, j+1} X_{j+1} +Y_j \tilde{Z}_{j, j+1}Y_{j+1} .\ee
   The gauge condition is $\tilde{X}_{j-1,j}(-1)^j Z_j \tilde{X}_{j, j+1} =1$ for all $j$. The disentangling transformation is shown in Fig.~\ref{fig:4}.
     
    For odd $j$,
\begin{align*}
        & Z_j \to \tilde{X}_{j-1,j}Z_j \tilde{Y}_{j, j+1}, \quad X_j \to X_j,  \\
        & \tilde{Z}_{j-1, j} \to -X_{j-1} \tilde{Z}_{j-1, j} X_{j}, \quad  \tilde{X}_{j-1,j} \to \tilde{X}_{j-1,j}. 
    \end{align*}
    For even $j$,
    \begin{align*}
        & Z_j \to - \tilde{Y}_{j-1,j}Z_j \tilde{X}_{j, j+1}, \quad X_j \to  \tilde{Y}_{j-1,j}Y_j \tilde{X}_{j, j+1},  \\
        & \tilde{Z}_{j-1, j} \to X_{j-1} \tilde{Z}_{j-1, j} X_{j}, \quad  \tilde{X}_{j-1,j} \to \tilde{Y}_{j-1,j}. 
    \end{align*}

\begin{figure}[tb]
    \centering
\includegraphics[width=0.9\linewidth]{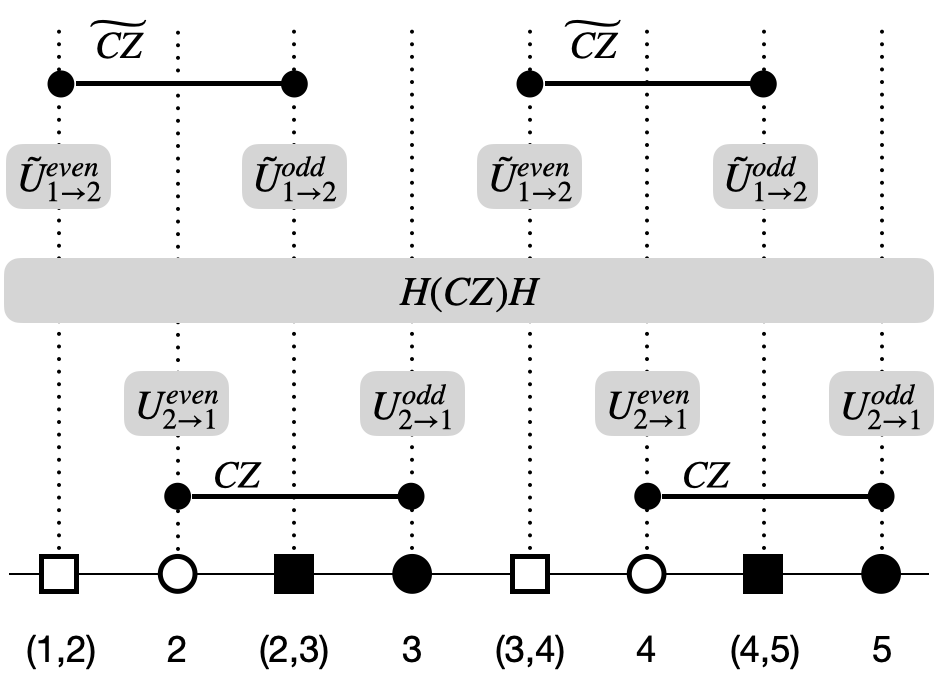}
    \caption{Self-dual disentangling transformation for $2\leftrightarrow 2$. Circled (squared) sites represent the original (dual) lattice. Filled (empty) marks represent odd (even) sites or bonds. Operators with tildes act on the dual lattice.}
    \label{fig:5}
\end{figure}

        \item $2 \leftrightarrow 2$ 
        
     The gauged Hamiltonian is 
    \begin{align}
        &  H' =  \sum_j X_{2j} \tilde{Z}_{2j, 2j+1} +\tilde{Z}_{2j, 2j+1} X_{2j+1} \\
        & -Z_{2j-1} \tilde{Z}_{2j-1, 2j}  X_{2j} Z_{2j+1}\nonumber \\
        & -Z_{2j}X_{2j+1} \tilde{Z}_{2j+1, 2j+2}  Z_{2j+2}.\nonumber
    \end{align} 
   The gauge conditions are \be \tilde{X}_{2j-1,2j} Y_{2j} \tilde{X}_{2j, 2j+1} Z_{2j+1} =1\ee and \be- Z_{2j} \tilde{X}_{2j, 2j+1} Y_{2j+1} \tilde{X}_{2j+1, 2j+2} =1\ee for all $j$.
   The disentangling transformation is shown in Fig.~\ref{fig:5}.
     
    For odd $j$,
    \begin{align*}
       & Z_j \to X_j, \quad X_j \to -X_{j-1} \tilde{Z}_{j-1, j} Y_j \tilde{Z}_{j,j+1},  \\
        & \tilde{Z}_{j-1, j} \to -\tilde{Z}_{j-2, j-1} X_{j-1} \tilde{Y}_{j-1, j} X_j, \quad  \tilde{X}_{j-1,j} \to \tilde{Z}_{j-1,j}. 
    \end{align*}
    For even $j$,
   \begin{align*}
       & Z_j \to X_j, \quad X_j \to \tilde{Z}_{j-1, j} Y_j \tilde{Z}_{j,j+1} X_{j+1} ,  \\
        & \tilde{Z}_{j-1, j} \to X_{j-1} 
 \tilde{Y}_{j-1, j} X_{j} \tilde{Z}_{j, j+1}, \quad  \tilde{X}_{j-1,j} \to \tilde{Z}_{j-1,j}. 
    \end{align*}

\begin{figure}
    \centering
\includegraphics[width=0.9\linewidth]{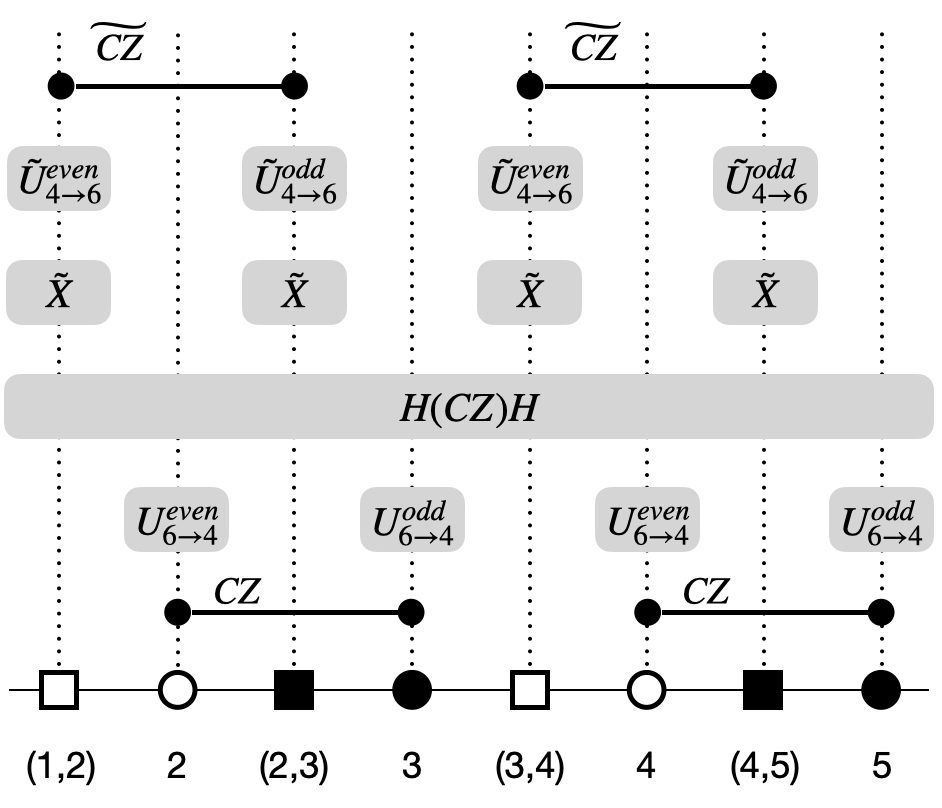}
    \caption{Self-dual disentangling transformation for $6\leftrightarrow 6$. Circled (squared) sites represent the original (dual) lattice. Filled (empty) marks represent odd (even) sites or bonds. Operators with tildes act on the dual lattice. }
    \label{fig:6}
\end{figure}

\item $6 \leftrightarrow 6$

The gauged Hamiltonian is 
    \begin{align}
        &  H' =  \sum_j X_{2j} \tilde{Z}_{2j, 2j+1} +\tilde{Z}_{2j, 2j+1} X_{2j+1} \\
        & +Z_{2j-1} \tilde{Z}_{2j-1, 2j}  X_{2j} Z_{2j+1} \nonumber\\
        & +Z_{2j}X_{2j+1} \tilde{Z}_{2j+1, 2j+2}  Z_{2j+2}.\nonumber
    \end{align} 
   The gauge conditions are \be -\tilde{X}_{2j-1,2j} Y_{2j} \tilde{X}_{2j, 2j+1} Z_{2j+1} =1\ee and \be- Z_{2j} \tilde{X}_{2j, 2j+1} Y_{2j+1} \tilde{X}_{2j+1, 2j+2} =1\ee for all $j$.
    The disentangling transformation is shown in Fig.~\ref{fig:6}.
     
    For odd $j$,
    \begin{align*}
       & Z_j \to X_j, \quad X_j \to -X_{j-1} \tilde{Z}_{j-1, j} Y_j \tilde{Z}_{j,j+1},  \\
        & \tilde{Z}_{j-1, j} \to \tilde{Z}_{j-2, j-1} X_{j-1} \tilde{Y}_{j-1, j} X_j, \quad  \tilde{X}_{j-1,j} \to \tilde{Z}_{j-1,j}. 
    \end{align*}
    For even $j$,
   \begin{align*}
       & Z_j \to -X_j, \quad X_j \to -\tilde{Z}_{j-1, j} Y_j \tilde{Z}_{j,j+1} X_{j+1} ,  \\
        & \tilde{Z}_{j-1, j} \to X_{j-1} 
 \tilde{Y}_{j-1, j} X_{j} \tilde{Z}_{j, j+1}, \quad  \tilde{X}_{j-1,j} \to -\tilde{Z}_{j-1,j}. 
    \end{align*}

\end{enumerate}

\subsection{Noninvertible symmetry transformations}
\label{append_noninv}
The noninvertible symmetry operators associated with the self-dualities of the $XY\pm YX$ models are directly related to the KW dual transformations. The corresponding noninvertible symmetry operators for the rest are obtained by unitary transformations.
In this section, we summarize their transformations in each model. 
\begin{enumerate}[label=(\alph*)]
    \item $1 \leftrightarrow 1$
    $$Z_j \to X_{j} X_{j+1}, \quad  X_j X_{j+1} \to  Z_{j+1}.$$
    \item $4 \leftrightarrow 4$
    $$Z_j \to X_{j} X_{j+1}, \quad  X_j X_{j+1} \to  -Z_{j+1}.$$
     \item $3 \leftrightarrow 3$
        $$Z_j \to \begin{cases} Y_jX_{j+1}, \quad &  \text{odd}\ j; \\
         X_j Y_{j+1}, \quad & \text{even}\ j.
        \end{cases}
        $$
        $$X_j Y_{j+1} \to \begin{cases} Y_j Z_{j+1} Y_{j+2}, \quad &  \text{odd}\ j; \\
         Z_{j+1}, \quad & \text{even}\ j.
        \end{cases}
        $$
        $$Y_j X_{j+1} \to \begin{cases}   Z_{j+1} , \quad &  \text{odd}\ j; \\
        X_j Z_{j+1}X_{j+2}, \quad & \text{even}\ j.
        \end{cases}
        $$
       \item $5 \leftrightarrow 5$
        $$Z_j \to \begin{cases}  X_j Y_{j+1}, \quad &  \text{odd}\ j; \\
       -Y_jX_{j+1}, \quad & \text{even}\ j.
        \end{cases}
        $$
        $$X_j Y_{j+1} \to \begin{cases} Z_{j+1} , \quad &  \text{odd}\ j; \\
         Y_j Z_{j+1} Y_{j+2} , \quad & \text{even}\ j.
        \end{cases}
        $$
        $$Y_j X_{j+1} \to \begin{cases} - X_j Z_{j+1}X_{j+2}  , \quad &  \text{odd}\ j; \\  -Z_{j+1} 
       , \quad & \text{even}\ j.
        \end{cases}
        $$
        
        \item $2 \leftrightarrow 2$
        $$X_j \to \begin{cases}  - Z_{j-1} X_{j} Z_{j+1}, \quad &  \text{odd}\ j; \\
        X_{j+1}, \quad & \text{even}\ j.
        \end{cases}
        $$
        $$Z_{j-1} X_j Z_{j+1} \to \begin{cases}  Z_{j}X_{j+1} Z_{j+2} , \quad &  \text{odd}\ j; \\
          -X_{j} , \quad & \text{even}\ j.
        \end{cases}
        $$
        $$Y_j Z_{j+1} \to \begin{cases} -Z_{j-1}X_j X_{j+1} Z_{j+2} , \quad &  \text{odd}\ j; \\  Z_j Z_{j+1} 
       , \quad & \text{even}\ j.
        \end{cases}
        $$
        (Note that for odd $j$, the gauge invariant operator is $\tilde{Z}_{j-1, j}Y_j \tilde{Z}_{j, j+1} Z_{j+1}$.) 
        $$Z_j Y_{j+1} \to \begin{cases} Y_{j+1} Y_{j+2}, \quad &  \text{odd}\ j; \\  -Z_{j+1} Z_{j+2} 
       , \quad & \text{even}\ j.
        \end{cases}
        $$
        (Note that for odd $j$, the gauge invariant operator is $Z_j \tilde{Z}_{j, j+1}  Y_{j+1} \tilde{Z}_{j+1, j+2}$.) 
         $$Z_j Z_{j+1} \to \begin{cases} Y_{j+1} Z_{j+2} , \quad &  \text{odd}\ j; \\  -Z_j Y_{j+1}
       , \quad & \text{even}\ j.
        \end{cases}
        $$

         \item $6 \leftrightarrow 6$
        $$X_j \to \begin{cases}   Z_{j-1} X_{j} Z_{j+1}, \quad &  \text{odd}\ j; \\
        X_{j+1}, \quad & \text{even}\ j.
        \end{cases}
        $$
        $$Z_{j-1} X_j Z_{j+1} \to \begin{cases}  Z_{j}X_{j+1} Z_{j+2} , \quad &  \text{odd}\ j; \\
          X_{j} , \quad & \text{even}\ j.
        \end{cases}
        $$
        $$Y_j Z_{j+1} \to \begin{cases} -Z_{j-1}X_j X_{j+1} Z_{j+2} , \quad &  \text{odd}\ j; \\  Z_j Z_{j+1} 
       , \quad & \text{even}\ j.
        \end{cases}
        $$
        (Note that for odd $j$, the gauge invariant operator is $\tilde{Z}_{j-1, j}Y_j \tilde{Z}_{j, j+1} Z_{j+1}$.) 
        $$Z_j Y_{j+1} \to \begin{cases} -Y_{j+1} Y_{j+2}, \quad &  \text{odd}\ j; \\  Z_{j+1} Z_{j+2} 
       , \quad & \text{even}\ j.
        \end{cases}
        $$
        (Note that for odd $j$, the gauge invariant operator is $Z_j \tilde{Z}_{j, j+1}  Y_{j+1} \tilde{Z}_{j+1, j+2}$.) 
         $$Z_j Z_{j+1} \to \begin{cases} Y_{j+1} Z_{j+2} , \quad &  \text{odd}\ j; \\  Z_j Y_{j+1}
       , \quad & \text{even}\ j.
        \end{cases}
        $$
\end{enumerate}

\section{Key elements of the compact boson CFT} 
\label{append:cft}
In this Appendix, we collect the essential elements in the compact boson CFT that are useful for our discussion in the main text. The reader can find more details in Refs.~\cite{ginsparg1988applied, francesco2012conformal}.

The compact boson CFT is described by the (Euclidean) action 
\be 
S = \frac{1}{4\pi} \int dz d\bar{z} \partial_z \phi \partial_{\bar{z}} \phi,
\ee
where $z = \exp(\tau +i x)$, and the free boson field $\phi$ is compactified on a circle of radius $R$, i.e., $\phi(z, \bar{z}) \sim \phi(z, \bar{z}) +2\pi R$. To reduce the periodicity to $2\pi$ , we simply rescale $\phi$ by $R$: $\phi \to \phi/R$. 
Split the rescaled $\phi(z, \bar{z})$ into the left-moving and the right-moving components: $\phi(z, \bar{z}) = X_L(z)+ X_R(\bar{z})$ and define $\theta(z, \bar{z}) = R^2[X_L(z)- X_R(\bar{z})]$.
At a generic radius, the CFT has global symmetry $U(1)^M \times U(1)^W$ generated by the momentum charge $\Qq^M$ and the winding charge $\Qq^W$, respectively, which act on $\phi$ and $\theta$ as :
\be 
 \phi  \to \phi +  \frac{1}{2}  \alpha_m,\quad \theta  \to \theta + \frac{1}{2} \alpha_w,
\ee
where  $\alpha_m \sim \alpha_m+2\pi $ and $\alpha_w \sim \alpha_w+2\pi $.

The local primary operators are:
\be 
V_{m, w}(z, \bar{z} )  = e^{ i m \phi(z, \bar{z}) + i w  \theta(z, \bar{z}) } , 
\ee
where $m \in  \bbz$ and $w\in \bbz$ are the momentum number and the winding number, i.e., the eigenvalues of $\Qq^M$ and $\Qq^W$, respectively. The conformal weights of $V_{m, w}$ are 
\be 
h_{m, w} = \frac{1}{4}  \left(\frac{m}{R} + w R\right) ^2, \quad \bar{h}_{m, w} = \frac{1}{4}  \left(\frac{m}{R} - w R\right) ^2,
\ee
and conformal dimensions are
\be \Delta_{m, w} =  h_{m, w} +\bar{h}_{m, w}= \frac{1}{2}(\frac{m^2}{R^2} + w^2 R^2).
\ee 
This is essentially Eq.~(\ref{eq:E}) in the main text:
\be 
E = \frac{1}{2} [\frac{(\Qq^M)^2}{R^2} + R^2 (\Qq^W)^2]. 
\label{eq:app_E}
\ee

Gauging the $\bbz_2$ subgroup of $U(1)^M$ maps $R \to R/2$, which is equivalent to $m \to 2 m $ and $w\to  w/2$. If we perform a T-duality transformation $m \leftrightarrow w$, then $\Delta_{m, w}$ is invariant at $R =\sqrt{2}$. This is the celebrated duality of the compact boson CFT at $R =\sqrt{2}$. 

If the compact boson CFT is twisted in space by $e^{i \pi \Qq^M}$, then there is a spectrum flow in $\Qq^W$. It now takes values in half-integers $w \in \bbz +\frac{1}{2}$ while $\Qq^M$ remains intact with $m \in \bbz$ \cite{cheng2023lieb}. In this case the conformal dimensions are 
\be \Delta_{m, w}  = \frac{1}{2}(\frac{m^2}{R^2} + w^2 R^2),
\ee 
which again can be expressed as Eq.~(\ref{eq:app_E}). $\Delta_{m,w} = \Delta_{m, -w}$ implies a double degeneracy in the spectrum for all $w$.  Again, if we perform a $\bbz_2$-gauging together with a T-duality transformation $\Qq^M \leftrightarrow \Qq^W$ such that $\Qq^M \to 2\Qq^W$ and $\Qq^W \to \Qq^M/2$, then the spectrum remains invariant at $R =\sqrt{2}$.

\section{General $L$} 
\label{append:odd}
In the main text, we restrict the lattice size $L$ to be even to avoid complications. On the one hand, for odd $L$, the models we consider are no longer unitarily equivalent, and whether there is a minus sign in front of the Hamiltonians can matter. On the other hand, some symmetry generators for even $L$ are no longer symmetry generators for odd $L$. For example, when $L$ is odd, even though the $XY+ YX$ model remains self-dual under gauging $\eta_Z =\prod_j Z_j$, 
$ \widetilde{Q}^{M} $  and $\widetilde{Q}^{W}$ are no longer conserved charges. 

As another example, the low energy states of the $XX$ model is four-fold degenerate for (large but finite) odd $L$. However, if we put a minus sign in front of its Hamiltonian, then the ground state of the model is two-fold degenerate.  These observations (which remain correct for the $XXZ$ model within a region of the parameter space) imply that in the IR, the $XX$ model must flow to a twisted compact boson CFT to allow degeneracies in the spectrum for odd $L$. Combining with the case for even $L$, we see that the degeneracy fits into the spectrum given by 
$ \Delta_{m, w}  = \frac{1}{2}(m^2/R^2 + w^2 R^2)$,
where eigenvalues $m$ and $w$ both take values in $\bbz  +L/2$. In other words, if we take the limit $L \to \infty$ with $L \mod 4$ fixed, then the $XX$ model flows to the compact boson CFT with $R =\sqrt{2}$ twisted by $C^L$ where $C =e^{i \pi(\Qq^M  +\Qq^W)}$ is also interpreted as an ``emanant" symmetry of the translation $T$ in Ref.~\cite{cheng2023lieb}.

Interestingly, for the $XY- YX$ model with odd $L$, even though the charge conjugation $\bbz_2^C$ symmetry is no longer well-defined, $Q^M$ and $Q^W$ remain conserved charges. Their quantized charges again depend on $L \mod 4$:
\be 
    m \in \bbz + \frac{L}{2}, \quad w \in \bbz + \frac{L}{4}. 
\ee 
Since the Hamiltonian can be mapped to free fermions, it is exactly solvable. Instead, we resort to small-scale numerics and conclude that the ground states (for odd $L >2$) are doubly degenerate with $m = \pm 1/2$ and $w = 1/4$ ($w = -1/4$) for $L =  1 \mod 4$ ($L =  3 \mod 4$).
 Note that even though $Q^M$ and $Q^W$ do not commute in general, they are commutative within the sub-Hilbert space of the ground states and are, therefore, simultaneously diagonalizable. In the limit $L \to \infty$, $Q^M$ and $Q^W$ become commutative \cite{pace2024lattice}. We can take the limit $L \to \infty$ while keeping $L \mod 4$ fixed.
Building upon our discussion of the $XY- YX$ model for even $L$ in the main text, we can unify the spectrum of the CFT for all $L$:  
$ \Delta_{m, w}  = \frac{1}{2}(m^2/R^2 + w^2 R^2)$, 
where $m$ takes values in $\bbz  +L/2 $ and $w$ takes values in $\bbz  +L/4 $. $R = \sqrt{2}$ remains fixed for all $L$. The ground state is two-fold degenerate for $L \neq 0 \mod 4$, in agreement with the spectrum on a lattice. Thus, for general $L$, the compact boson CFT is  twisted by $e^{i\pi 
 L(\frac{1}{2} \Qq^M + \Qq^W )}$. This should be compared with the case for the Levin-Gu model where $\Qq^M$ takes values in $\bbz$ and $\Qq^W$ takes values in $\bbz  +L/4 $. The compact boson CFT for the Levin-Gu model is only twisted by $e^{i\pi L\frac{1}{2} \Qq^M}$ \cite{cheng2023lieb} \footnote{$\Qq^M$ and $\Qq^W$ are switched in Ref.~\cite{cheng2023lieb}.} .

Let us return to the self-duality of the $XY-YX$ model for a general $L$. Similar to the discussion in the main text, the model is self-dual under gauging $\bbz_2^{\eta_Z}$. For general $L$, gauging $\bbz_2^{\eta_Z}$ maps $Q^M \to 2Q^W$ and $Q^W \to Q^M/2$. It projects out states with $m = (2s+1)+L/2$ for $s\in \bbz$ and adds states with $w = s/2+L/4$ for $s\in \bbz$. This is again compatible with the continuum spectrum $\Delta_{m, w}$ which is invariant under the transformations $\Qq^M \to 2 \Qq^W$ and  $\Qq^W \to \Qq^M/2$ at $R =\sqrt{2}$.

\bibliography{XXModel} 

\end{document}